\documentclass[twocolumn,aps,prl,superscriptaddress]{revtex4-1}
\usepackage{graphicx,subfigure}
\usepackage{amsmath,amssymb,bm}
\usepackage{mathtools}
\usepackage{multirow}
\usepackage{makecell}
\usepackage{textcomp}
\usepackage{float}
\usepackage{xcolor, soul}

\usepackage{color}
\usepackage[normalem]{ulem}
\usepackage{dsfont}
\usepackage{mathrsfs}
\usepackage{braket}
\usepackage{transparent}
\usepackage{xcolor}
\usepackage{hhline}
\usepackage{graphicx}
\usepackage{pdfpages}
\usepackage{bibunits}
\usepackage{filecontents,hyperref}

\makeatletter
\AtBeginDocument{\let\LS@rot\@undefined}
\makeatother

\usepackage{graphicx}

\graphicspath{ {figures/} }

\makeatletter
\let\saved@includegraphics\includegraphics
\AtBeginDocument{\let\includegraphics\saved@includegraphics}
\makeatother

\newcommand{\figOne}{
 \begin{figure*}[t]
    \centering
    \includegraphics[width=\textwidth]{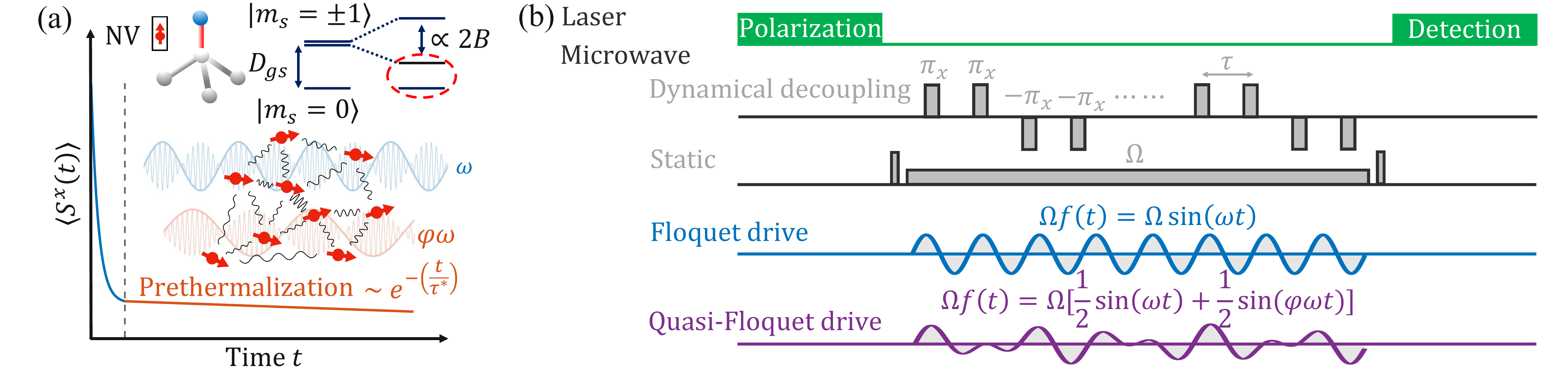}
    \caption{{\bf Quasi-Floquet prethermalization in a dipolar spin system.} (a) Quasi-periodic driving of a strongly interacting NV ensemble in diamond with two incommensurate frequencies $\omega$ and $\varphi \omega$, where $\varphi$ is the golden ratio. Typical thermalization dynamics of the spin polarization $\langle S^x(t)\rangle$ exhibit an initial fast decay followed by a late-time slow relaxation. Top: Level structure of NV center. Without an external field, $|m_s = \pm1\rangle$ sublevels are degenerate and sit $D_{gs}=(2\pi)\times2.87~$GHz above $|m_s=0\rangle$. A magnetic field $B\sim350~$G along the NV axis splits $|m_s = \pm1\rangle$, enabling the isolation of a two-level system. (b) Experimental sequence. 
    The dynamical decoupling sequence eliminates the on-site random fields induced by the environmental bath spins. The sequence includes a series of fast $\pi$-pulses with alternating phases along $\hat{x}$ and $-\hat{x}$ axes to compensate the pulse errors. The inter-pulse spacing is fixed at $\tau = 0.1~\mu$s, much smaller than the interaction timescale between NV centers. A static microwave $\Omega \sum_{i} S^x_i$ together with dipolar interaction serves as the static Hamiltonian $\mathcal{H}_0$, and a time-dependent microwave $\Omega f(t) \sum_{i} S^x_i$ serves as the Floquet and quasi-Floquet drives. A final $\frac{\pi}{2}$-pulse along $\mp \hat{y}$ axis rotates the spins back to $\hat{z}$ for detection \cite{SM}.
    }
    \label{fig:fig1}
\end{figure*}
}

\newcommand{\figTwo}{
  \begin{figure*}[t]
    \centering
    \includegraphics[width=0.95\textwidth]{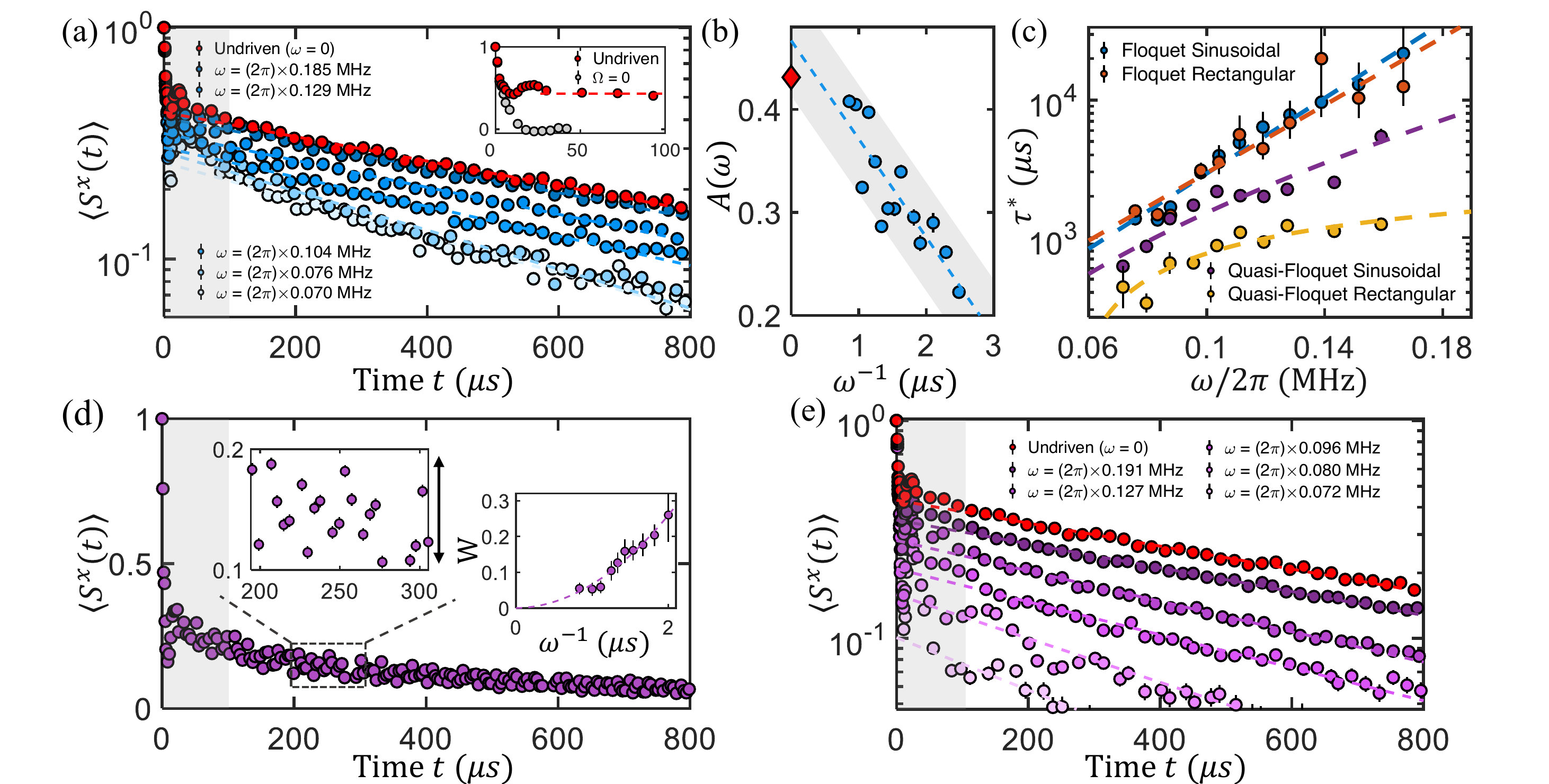}
    \caption{{\bf Probing the spin dynamics under periodic and quasi-periodic drives.} (a) The measured spin polarization $\langle S^x (t)\rangle$ under Floquet drive. After an initial fast decay (gray shaded area), a long-lived prethremal regime persists. Dashed lines are fits of the late-time dynamics using single exponential decay.
    Inset: Measured initial spin dynamics for the undriven case (red): After the initial fast relaxation, the polarization decays to an equilibration plateau (dashed line), due to the finite static field, $\Omega \sum_{i} S^x_i$. In contrast, the polarization with only dynamical decoupling ($\Omega = 0$) quickly decays to zero. (b) Measured prethermal equilibrium value, $A(\omega)$, as a function of $\omega^{-1}$. Dashed line is a linear fit with the gray shaded area representing 95\% confidence interval. The red diamond marks the measured amplitude for the undriven case. (c) Heating timescale $\tau^{*}$ as a function of the driving frequency $\omega$. For both 
    sinusoidal and rectangular single-frequency drives, $\tau^{*}\sim \mathcal{O}(e^{\omega/J})$. For 
    sinusoidal quasi-periodic drive, $\tau^{*}\sim \mathcal{O}(e^{\omega^{\frac{1}{2}}})$; while for rectangular quasi-periodic drive, $\tau^{*}\sim \mathcal{O}(\omega^{\frac{1}{2}})$.
    (d) Measured spin dynamics under quasi-periodic drive [$\omega = (2\pi)\times0.103~\mathrm{MHz}$]. 
    We observe an additional small time-quasiperiodic micromotion on top of a slow relaxation. Left inset: zoom-in of the micromotion. Right inset: relative amplitude of the micromotion, $W$, scales quadratically with $\omega^{-1}$. (e) Using rolling-average to remove the micromotion \cite{SM}, we observe a quasi-Floquet prethermal regime, whose lifetime $\tau^{*}$ increases with $\omega$. Dashed lines are fits using single exponential decay. %
    Errorbars on the spin polarization represent 1 s.d. accounting statistical uncertainties, and errorbars on the prethermal plateau and timescale represent 1 s.d. from the fitting.
    }
    \label{fig:fig2}
\end{figure*}
}

\newcommand{\figThree}{
\begin{figure*}[t!]
    \centering
    \includegraphics[width=\textwidth]{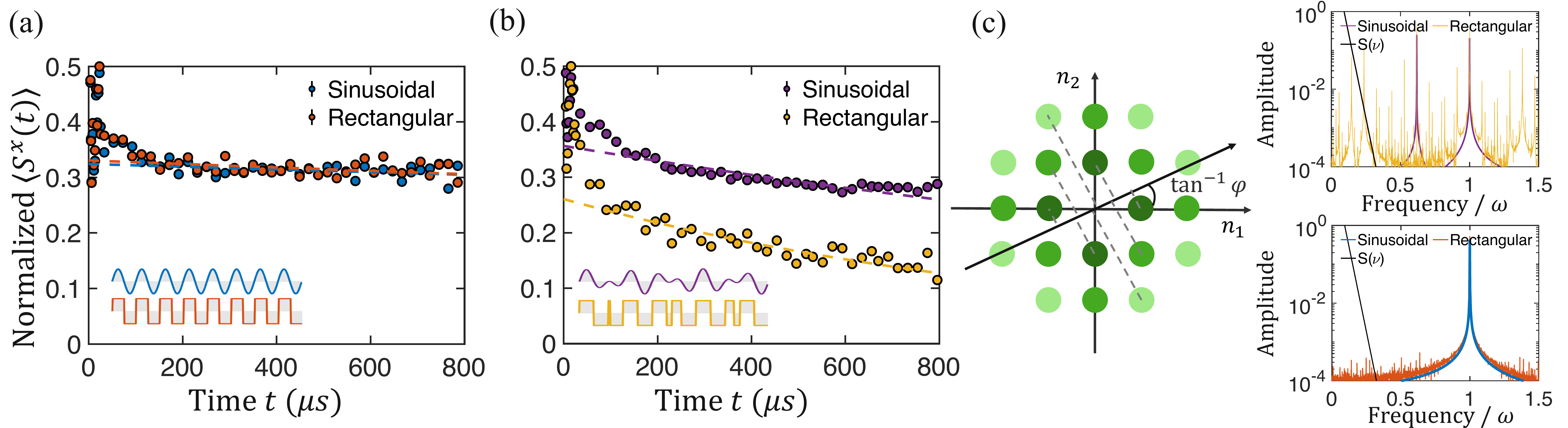}
    \caption{{\bf Comparison between sinusoidal and rectangular driving fields.} (a) A representative late-time dynamics of the spin ensemble under single-frequency (Floquet) drive [$\omega = (2\pi)\times0.151~\mathrm{MHz}$]. 
    Here the plotted polarization is normalized by the intrinsic decay timescale, $T_0$, measured without the driving field, i.e. normalized $\langle S^\mathrm{x}(t)\rangle\sim e^{-t/\tau^*}$. 
    (b) Under quasi-periodic drive [$\omega = (2\pi)\times0.143~\mathrm{MHz}$], the spin dynamics is extremely sensitive to the smoothness of the drive, and displays a significantly faster decay under a rectangular drive.
    Errorbars represent 1 s.d. accounting statistical uncertainties. 
    (c) Two-dimensional Fourier spectrum $F_{n_1,n_2}$ of the driving field $f(t)$ \cite{SM}. The horizontal (vertical) axis corresponds to the absorption of $n_1$ ($n_2$) photons with energy $\omega$ ($\varphi\omega$) from the drive, where $n_1, n_2 = 0, \pm1, ...$. The color of each green dot represents the amplitude of $F_{n_1,n_2}$ at frequency $\nu = n_1 \omega + n_2\varphi\omega$ (darker represents larger amplitude). The Fourier spectrum $F(\nu)$ can be thought of as a projection of green dots onto a line with slope $\varphi$. Right panel: Fourier spectrum $F(\nu)$ of quasi-periodic and periodic drive with sinusoidal and rectangular amplitude. For the quasi-Floquet scenario, the spectrum with rectangular drive exhibits resonances at the small frequency range, which overlaps with the system's local spectral function $S(\nu)$ (black) and leads to energy absorption. 
    }
    \label{fig:fig3}
\end{figure*}
}

\makeatletter
\newcommand*{\centerfloat}{%
  \parindent \z@
  \leftskip \z@ \@plus 1fil \@minus \textwidth
  \rightskip\leftskip
  \parfillskip \z@skip}
\makeatother

\usepackage{lineno} 
\begin{document}

\title{Quasi-Floquet prethermalization in a disordered dipolar spin ensemble in diamond}

\author{
Guanghui~He,$^{1,*}$
Bingtian~Ye,$^{2,3,*,\dag}$
Ruotian~Gong,$^{1}$ 
Zhongyuan~Liu,$^{1}$ \\
Kater W. Murch,$^{1,4}$
Norman~Y.~Yao,$^{2,3,5}$
Chong~Zu,$^{1,4,\ddag}$
\\
\normalsize{$^{1}$Department of Physics, Washington University, St. Louis, MO 63130, USA}\\
\normalsize{$^{2}$Department of Physics, Harvard University, Cambridge, MA 02138, USA}\\
\normalsize{$^{3}$Department of Physics, University of California, Berkeley, CA 94720, USA}\\
\normalsize{$^{4}$Institute of Materials Science and Engineering, Washington University, St. Louis, MO 63130, USA}\\
\normalsize{$^{5}$Materials Science Division, Lawrence Berkeley National Laboratory, Berkeley, CA 94720, USA}\\
\normalsize{$^*$These authors contributed equally to this work.}\\
\normalsize{$^\dag$To whom correspondence should be addressed; E-mail:  bingtian\_ye@berkeley.edu}\\
\normalsize{$^\ddag$To whom correspondence should be addressed; E-mail:  zu@wustl.edu}\\
}

\begin{abstract}
Floquet (periodic) driving has recently emerged as a powerful technique for engineering quantum systems and realizing  non-equilibrium phases of matter \cite{eckardt2017colloquium,oka2019floquet,harper2020topology, moessner2017equilibration, bukov2015universal,  Zhang2017DTC,Choi2017timecrystal,Choi2020HEngineer,Geier2021HEngineer, mi2022time, yao2017discrete, else2016floquet, khemani2016phase, machado2020long, potirniche2017floquet,  Eisert2015NonEquilibrium, potirniche2017floquet, ye2021floquet,kyprianidis2021observation, randall2021many, rechtsman2013photonic, dumitrescu2018logarithmically, long2021nonadiabatic}. 
A central challenge to stabilizing quantum phenomena in such systems is the need to prevent  energy absorption from the driving field.
Fortunately, when the frequency of the drive is  significantly larger than the local energy scales of the many-body system, energy absorption is suppressed \cite{Peng2021Prethermal,rubio2020floquet,beatrez2021floquet,abanin2017rigorous, weidinger2017floquet, d2014long, Ho22quantum,Ye2020hydro,Singh2019PRX,Viebahn2021PRX}.
The existence of this so-called prethermal regime depends sensitively on the range of interactions and the presence of multiple driving frequencies
\cite{ho2018bounds, machado2019exponentially,machado2020long, Else2020quasi, mori2021rigorous}.
Here, we report the observation of Floquet prethermalization in a strongly interacting dipolar spin ensemble in diamond, where the angular dependence of the dipolar coupling helps to mitigate the long-ranged nature of the interaction.
Moreover, we extend our experimental observation to quasi-Floquet drives with \emph{multiple} incommensurate frequencies.
In contrast to a \emph{single}-frequency drive, we find that the existence of prethermalization is extremely sensitive to the smoothness of the applied field. 
Our results open the door to stabilizing and characterizing non-equilibrium phenomena in quasi-periodically driven systems.

\end{abstract}
\date{\today}
\maketitle

Floquet theory describes the dynamics of a system whose Hamiltonian exhibits a \emph{single} time-translation symmetry.
Often used as a tool to control quantum systems, Floquet engineering (i.e.~periodic driving) can help to prevent environment-induced decoherence and more recently, has enabled the study of novel quantum dynamical phenomena~\cite{hahn1950spin, waugh1968approach, viola1999dynamical, Choi2020HEngineer, zhou2020quantum, gong2022coherent, hahn2021long}.
In particular, many-body Floquet systems can host intrinsically non-equilibrium phases of matter, ranging from discrete time crystals \cite{Zhang2017DTC,Choi2017timecrystal, mi2022time, kyprianidis2021observation, randall2021many, yao2017discrete, else2016floquet, rovny2018observation, beatrez2023critical} to Floquet topological states \cite{lindner2011floquet, rechtsman2013photonic, Wang2013TopologicalInsulator, potirniche2017floquet,po2016chiral, po2017radical, zhang2022digital, dumitrescu2022dynamical, long2021nonadiabatic, xu2022topological}.
Even richer non-equilibrium behaviors can arise in ``quasi-Floquet" systems, where a \emph{single} time-translation symmetry is replaced by \emph{multiple} time-translation symmetries \cite{Else2020quasi, qi2021universal, peng2018time, blekher1992floquet, dumitrescu2018logarithmically, zhao2021random}. 
For instance, the spontaneous breaking of the latter can result in time quasi-crystalline order, which features a  subharmonic response that is fundamentally distinct from conventional time crystals~\cite{dumitrescu2018logarithmically, Else2020quasi}.

A critical obstacle to stabilizing and observing such phenomena in driven quantum systems is Floquet heating: the inevitable absorption of  energy from the driving field.
One potential solution arises when the driving frequency, $\omega$, is significantly larger than the local energy scale, $J$, of the many-body system; in this case, Floquet prethermalization occurs and there exists an exponentially long-lived preheating regime described by an effective time-independent Hamiltonian, $\mathcal{H}_\mathrm{eff}$~\cite{Peng2021Prethermal,rubio2020floquet,beatrez2021floquet,abanin2017rigorous, weidinger2017floquet, d2014long, Ho22quantum,Ye2020hydro}.
The intuition underlying Floquet prethermalization is simple --- in order to absorb a single photon from the drive, the system must undergo $\sim \omega/J$ off-resonant rearrangements of its local degrees of freedom.
This higher-order process leads to an exponentially slow heating rate $\sim \mathcal{O}(e^{-\omega/J})$.

Despite this promise, there are two natural scenarios where prethermalization can break down: (i) systems with long-range, power-law interactions~\cite{ho2018bounds, machado2020long} and (ii)  quasi-Floquet systems where multi-photon processes can enable resonant energy absorption \cite{Else2020quasi}. 

\figOne

Here, we report the experimental observation of Floquet prethermalization in a long-range interacting quantum system under quasi-periodic driving.
Our experimental platform consists of a dense ensemble of dipolar interacting nitrogen-vacancy (NV) centers in diamond (Fig.~\ref{fig:fig1}) \cite{doherty2013nitrogen, Choi2017timecrystal, Choi2020HEngineer, zhou2020quantum,  Zu2021emergent, davis2021probing}.
With single-frequency modulation, we observe that the heating time, $\tau^{*}$, is consistent with an exponential scaling with increasing driving frequency.
%
In contrast, by driving quasi-periodically with two frequencies, we find that Floquet heating can be fitted to a stretched exponential profile with $\tau^{*}\sim \mathcal{O}(e^{\omega^\frac{1}{2}})$~\cite{Else2020quasi}.
%
Interestingly, in the quasi-periodic case, the heating is extremely sensitive to the \emph{smoothness} of the drive; indeed, when the system is driven via rectangular pulses (as opposed to sinusoidal pulses), we observe a significant enhancement in the heating rate (Fig.~\ref{fig:fig3})~\cite{Else2020quasi}.
We remark that the presence of slow heating is reminiscent of classic results from the NMR literature detailing the observation of long-lived dynamics in driven systems~\cite{Maricq1987SpinT, Maricq1990LongTimeLO, Sakellariou1999QuasiEqu, waugh1968approach, carravetta2004long, redfield1969nuclear}.
However, the origin of the long-lived dynamics are either expected to be independent of the driving frequency or exhibits a lifetime that scales as a power-law of the driving frequency \cite{SM}. 
These scenarios are markedly distinct from the context of Floquet prethermalization, which exhibits an exponentially long lifetime in $\omega$.
Despite such distinction, how to unequivocally demonstrate exponential over power-law scaling from real experiment remains a challenging task \cite{SM}.

\emph{Experimental system}---We choose to work with a diamond sample containing a dense ensemble of spin-1 NV centers with concentration, $\rho \sim 4.5~$ppm~\cite{SM, doherty2013nitrogen}. 
The NV centers can be optically initialized 
and read out using green laser. 
In the presence of an external magnetic field $\sim 350~$G, the $|m_s = \pm 1 \rangle$ sublevels are Zeeman split, allowing us to isolate an effective two-level system, $\{|m_s = 0\rangle, |m_s = -1\rangle\}$ (Fig.~1a).
By applying a resonant microwave field with Rabi frequency $\Omega$, the effective  Hamiltonian governing the system (in the rotating frame) is~\cite{Choi2020HEngineer, Choi2017timecrystal, Zu2021emergent}:
\begin{equation} \label{eq1}
\begin{split}
\mathcal{H}_{0} = -\sum_{i<j} \frac{J_0 \mathcal{A}_{i,j}}{r^3_{i,j}}(S^z_i S^z_j - S^x_i S^x_j - S^y_i S^y_j) + \Omega \sum_{i} S^x_i,
\end{split}
\end{equation}
where $J_0 = (2\pi)\times52~$MHz$\cdot$nm$^3$,  $\mathcal{A}_{i,j}$ characterizes the  angular dependence of the dipolar interaction, $r_{i,j}$ is the distance between the $i^{th}$ and $j^{th}$ NV centers, and $\hat{S}$ is spin operator. 

\figTwo

We note that $\mathcal{H}_{0}$  contains only the energy-conserving terms of the dipolar interaction under the rotating-wave approximation.
The approximation holds because the NV transition frequency
is more than three orders of magnitude larger than any other terms in the interacting Hamiltonian \cite{SM}. 
The presence of other paramagnetic spins in the diamond lattice, such as $^{13}$C nuclear spins and substitutional nitrogen impurities, leads to an additional on-site random field at each NV center which is eliminated using dynamical decoupling (Fig.~1b) \cite{SM}. 

Let us begin by characterizing the dynamics of the NV ensemble under the static Hamiltonian $\mathcal{H}_0$.
We set $\Omega = (2\pi)\times0.05~$MHz, comparable to the average dipolar interaction strength.
After optically initializing the NV spins to $|m_s=0\rangle$, we then prepare a product state, $\otimes_{i} \frac{|0\rangle_i+|-1\rangle_i}{\sqrt{2}}$, by applying a global $\pi/2$-pulse around the $\hat{y}$ axis.
We let the system evolve  under $\mathcal{H}_0$ for a time $t$, before measuring the final NV polarization, $\langle S^x(t)\rangle$, along the $\hat{x}$ direction.

The polarization dynamics proceed in two steps. 
At early times, $t\lesssim 100~\mu$s, the polarization exhibits rapid decay toward a plateau value, reflecting  local equilibration under $\mathcal{H}_0$ (Fig.~\ref{fig:fig2}a).
Following these initial dynamics, the system exhibits a slow exponential decay $\sim A_0 e^{-\frac{t}{T_0}}$ with $A_0 = (0.43\pm0.01)$ and a time-scale $T_0 = (0.82\pm0.03)~$ms that is  consistent with spin-phonon relaxation \cite{jarmola2012temperature}. To ensure that the observed spin dynamics does not come from the incorporated dynamical decoupling pulses, we also investigate the corresponding spin dynamics at $\Omega = 0$. The measured NV polarization quickly decays to zero, in agreement with the expectation of thermalization behavior (Fig.~2a Inset).

\emph{Floquet prethermalization}---To probe the nature and existence of Floquet prethermalization, we modulate the Rabi frequency as  $\Omega(t) = \Omega [1+f(t)]$, where $f(t) = \mathrm{sin}(\omega t)$  (Fig.~\ref{fig:fig1}b). We note that $\Omega(t)$ contains two fundamentally different components: the constant field, $\Omega \sum_{i} S^x_i$, is a part of the previous undriven Hamiltonian $\mathcal{H}_0$, while the time-dependent component $\Omega f(t) \sum_{i} S^x_i$ acts as the Floquet drive.
Starting with a driving frequency $\omega = (2\pi)\times 0.07~\mathrm{MHz}$, which is comparable to energy scales within $\mathcal{H}_0$, we perform the same spin polarization measurement (light blue curve in Fig.~\ref{fig:fig2}a).
The measured spin dynamics at stroboscopic times, $t = 2\pi N/\omega$ (where $N$ is an integer), exhibit an initial relaxation, which is qualitatively similar to the undriven case.
However, the late-time dynamics exhibit a significantly faster polarization decay, arising from Floquet heating. 
To obtain the heating timescale $\tau^{*}$, we fit the experimental data to $\sim A e^{-(\frac{t}{\tau^*}+\frac{t}{T_0})}$, where $T_0$ is the previously  extracted spin-phonon lifetime. 
As shown in Fig.~\ref{fig:fig2}a, by increasing the driving frequency, one can extract the frequency dependence of the amplitude, $A(\omega)$, and the heating time-scale, $\tau^* (\omega)$; both are crucial for understanding the nature of Floquet prethermalization.  

Focusing first on the heating time-scale, we find that $\tau^{*}$ increases exponentially with $\omega$ for more than an order of magnitude, demonstrating the existence of a parametrically long-lived prethermal regime (Fig.~\ref{fig:fig2}c).
%
%
The observed exponential scaling also allows us to extract a phenomenological local energy scale of the NV many-body system, $J_\mathrm{exp} = (2\pi)\times(0.032\pm0.006)$~MHz. 

Intuitively, this $J_\mathrm{exp}$ extracted from Floquet heating process is expected to agree with the local energy scale of the system.
However, for systems with power-law interaction as ours ($\sim 1/r^3$ in 3D), a naive estimation of the local energy scale $J \approx \int \frac{J_0}{r^3} \rho  d^3 \textbf{r} $ is divergent, and thus, one should expect the prethermalization to not exist. 
Nevertheless, an important missing piece to this puzzle is the angular dependence of the dipolar interaction, $\mathcal{A}_{i,j}$~\cite{ho2018bounds}.
Crucially, the combination of this angular dependence and the NV's random positioning in the diamond lattice ensures that the average, $\overline{\mathcal{A}}_{i,j} = 0$, which helps to mitigate the divergence of the above integral. We note that even if the NV centers are located on a three-dimensional regular lattice, we expect the angular dependence should also average to zero over a large lengthscale, so the positional disorder is not necessary here.

A more careful analysis reveals that the relevant local energy scale is the variance of the interaction,  $\widetilde{J}  \approx  [\int (\frac{J_0 \mathcal{A}_{i,j}}{r^3})^2 \rho d^3 \textbf{r}]^{\frac{1}{2}} = \sqrt{\frac{16\pi}{15}} J_0 \rho$ \cite{SM}.
As long as the driving frequency $\omega > \widetilde{J}$, one should expect the presence of prethermalization,
in agreement with previous theoretical studies \cite{ho2018bounds}.
Using the independently calibrated NV density, $\rho$,
we estimate the local energy scale $\widetilde{J} \approx (2\pi)\times 0.02$~MHz, which is indeed comparable with $J_\mathrm{exp}$ extracted from Floquet heating.

Next, let us turn to analyzing the frequency dependence of the amplitude, $A(\omega)$.
%
One can think of $A(\omega)$ as the value of the prethermal plateau. 
In general, for short-range interactions, it is expected that $A(\omega)$ is determined by a time-independent effective Hamiltonian, $\mathcal{H}_\mathrm{eff}(\omega) = \mathcal{H}_0 + \mathcal{O}(\frac{\widetilde{J}}{\omega})$, which can be calculated order-by-order via a Magnus expansion~\cite{kuwahara2016floquet}. 
In this case,  $A(\omega) = \mathrm{Tr}[\sum_{i}S_i^x e^{-\beta H_\mathrm{eff}(\omega)}] = A_0 + \mathcal{O}(\frac{\widetilde{J}}{\omega})$, where the inverse temperature $\beta$, is set by the energy density of the initial state.

For sufficiently long-range interactions (such as dipolar interactions in 3D), the existence of $\mathcal{H}_\mathrm{eff}$ is unproven \cite{else2017prethermal, kuwahara2016floquet, mori2016rigorous, abanin2017rigorous, machado2020long}.
However, by probing the functional form of $A(\omega)$ and its extrapolated value as $\omega \rightarrow \infty$, one can provide experimental evidence for the existence of $\mathcal{H}_\mathrm{eff}$.
As depicted in Fig.~\ref{fig:fig2}b, we find that the frequency dependence of $A(\omega)$ is linear in $\omega^{-1}$, allowing us to extrapolate $A(\omega \rightarrow \infty) = (0.47\pm0.06)$. 
This is consistent with the measured value in the undriven case, $A_0 = (0.43\pm0.01)$, suggesting that despite the presence of strong
long-range interactions, the effective Hamiltonian exists and can be well-approximated by $\mathcal{H}_0$ at leading order \cite{SM}.

\emph{Quasi-Floquet prethermalization}---
We now turn to the quasi-Floquet setting.
Specifically, we choose $f(t) = \frac{1}{2}[\mathrm{sin}(\omega t)+\mathrm{sin}(\varphi\omega t)]$,
where $\varphi = (\sqrt{5}-1)/2\approx 0.618$ is the golden ratio, so that the system is driven by two incommensurate frequencies.
From the perspective of Floquet heating, the situation is significantly more complex.
In particular, recall that  within Fermi's golden rule, the heating rate can be estimated from the overlap between the Fourier spectrum of the drive, $F(\nu) = \int f(t)e^{i\nu t} dt$, and the local spectral function of the spin ensemble, 
$S(\nu) = \sum_{i,j} \delta(E_i-E_j-\nu) |\langle i |S^x|j\rangle|^2$, where $E_i$ and $|i\rangle$ are the eigenenergy and eigenstate of the spin system.

This picture immediately provides a  more formal intuition for the exponentially slow heating observed in the context of  single-frequency driving (Fig.~\ref{fig:fig3}c).
In particular, for $f(t) = \mathrm{sin}(\omega t)$, $F(\nu)$ exhibits a cut-off at  frequency $\omega$.
Meanwhile, as aforementioned, $S(\nu)$ exhibits an exponentially small tail for  frequencies $\nu>\widetilde{J}$. This is specifically for our case where we need the dipolar interaction to average to zero, in general this is true for the local energy scale J.
In combination, this implies that for a single driving frequency,  $\omega>\widetilde{J}$, energy absorption is strongly suppressed leading to $\tau^* \sim e^{\omega/\tilde{J}}$.

For driving with two incommensurate frequencies, even when $\omega>\widetilde{J}$, there are multi-photon processes that are effectively resonant within the local energy scale; these processes correspond, for example, to the absorption of $n_1$ photons of energy $\omega$ and the emission of $n_2$ photons of energy $\varphi \omega$.
Thus, there is no strict cut-off for $F(\nu)$, and the drive spectrum exhibits a non-zero amplitude for all frequencies $\nu = |n_1 \omega - n_2\varphi \omega| < \widetilde{J}$.
Interestingly, despite this, for sufficiently large driving frequencies, seminal results have proven that the quasi-Floquet heating timescale remains extremely slow, exhibiting a stretched exponential lower-bound. The predicted lower bound is $\mathcal{O}(e^{C(\omega/\widetilde{J})^{\frac{1}{m}}})$, where $m$ is the number of incommensurate driving frequencies and $C$ is a dimensionless factor \cite{Else2020quasi}.

\figThree

In contrast to the Floquet case, we measure the dynamics at evenly spaced time points, since there does not exist a stroboscopic time which is an integer multiple of \emph{both} drives.
Much like the single-frequency drive, after an early-time transient, the spin polarization exhibits a slow decay.
However, we observe small oscillations scaling as $\sim \omega^{-2}$ on top of the decay (Fig.~\ref{fig:fig2}d), corresponding to the micromotion of the quasi-Floquet system \cite{sambe1973steady, goldman2014periodically, eckardt2015high}. We note that for a single-frequency drive, similar micromotion will also emerge if one does not measure the spin dynamics at stroboscopic times \cite{SM}. Intuitively, such micromotion arises from the time-dependent portion of the Hamiltonian which only averages to zero for each complete Floquet cycle.
To reliably extract a heating timescale $\tau^{*}$ from our quasi-Floquet measurements, we perform a rolling average to obtain the overall decay profile  (Fig.~\ref{fig:fig2}e) \cite{SM}.
By varying the driving frequency,  we extract a heating time-scale, $\tau^{*} \sim e^{\omega^{\frac{1}{2}}}$ (Fig.~\ref{fig:fig2}c), which is  consistent with the theoretically predicted stretched exponential form 
\cite{Else2020quasi}.

\emph{Robustness of quasi-Floquet prethermalization}---The stability of slow prethermal heating is quite different depending on whether one considers the Floquet or quasi-Floquet setting. 
For the Floquet setting, the exponential behavior of $\tau^*$ is robust to the functional form of the drive amplitude $f(t)$.
However, in the quasi-Floquet setting, the stretched exponential behavior of $\tau^*$ is \emph{only} predicted to hold when $f(t)$ is smooth.
In particular, when $f(t)$ is smooth, even though $F(\nu)$ does not exhibit a cut-off for small $\nu$, its amplitude is exponentially small in this regime (Fig.~\ref{fig:fig3}c) \cite{Else2020quasi, mori2021rigorous}.

These expectations are indeed borne out by the data (Fig.~\ref{fig:fig2}c, Fig.~\ref{fig:fig3}).
Using a rectangular wave $f(t) = \mathrm{Sgn}[\frac{1}{2}\mathrm{sin}(\omega t)+\frac{1}{2}\mathrm{sin}(\varphi\omega t)]$, we observe
that the heating timescale is significantly shortened and scales as a power-law with increasing driving frequency $\sim\omega^{\frac{1}{2}}$, as opposed to a stretched exponential. 
In contrast, for a single-frequency drive, the smoothness of the driving field is not critical: the Floquet heating time-scale exhibits an exponential scaling for both sinusoidal and rectangular forms of $f(t)$.
%

\emph{Outlook}---Looking forward, our work opens the door to a number of intriguing future directions.
First, it is interesting to ask whether the restriction on $\overline{\mathcal{A}}_{i,j} = 0$  
is essential for realizing prethermalization in long-range interacting systems
 \cite{ho2018bounds, machado2019exponentially, machado2020long}.
%
%
Second, the observed long-lived quasi-Floquet prethermal regime can enable the experimental realization of novel non-equilibrium phases of matter \cite{lindner2011floquet, rechtsman2013photonic, Wang2013TopologicalInsulator, potirniche2017floquet,po2016chiral, po2017radical, zhang2022digital, dumitrescu2022dynamical, long2021nonadiabatic, xu2022topological}.
Finally, while our experiments suggest the presence of power-law-slow-heating in the case of a quasi-Floquet, rectangular-wave drive, the precise frequency dependence of the heating rate remains unknown and requires future study. 

\smallskip

\emph{Acknowledgements:} We gratefully acknowledge the insights of and discussions with E. Henriksen, P.~Y. Hou, A. Jayich, F. Machado, P. Peng, G. Refael, and W. Wu. We thank C. Gaikwad, D. Kowsari, and K. Zheng for their assistance in setting up the experiment. This work is supported by the Startup Fund, the Center for Quantum Leaps, the Institute of Materials Science and Engineering and the OVCR Seed Grant from Washington University. B.Y. acknowledges support from the U.S. Department of Energy (BES grant No. DE-SC0019241). K.W.M. acknowledges support from NSF Grant No. PHY-1752844 (CAREER). N.Y.Y. acknowledges support from the U.S. Department of Energy, Office of Science, through the Quantum Systems Accelerator (QSA), a National Quantum Information Science Research Center and the David and Lucile Packard foundation. 


%
%


\bibliographystyle{naturemag}
\bibliography{Main.bib}

\end{document}


\title{Supplementary Material:\\
Quasi-Floquet prethermalization in a disordered dipolar spin ensemble in diamond}

\author{
Guanghui~He,$^{1,*}$
Bingtian~Ye,$^{2,3,*,\dag}$
Ruotian~Gong,$^{1}$ 
Zhongyuan~Liu,$^{1}$ \\
Kater W. Murch,$^{1,4}$
Norman~Y.~Yao,$^{2,3,5}$
Chong~Zu,$^{1,4,\ddag}$
\\
\normalsize{$^{1}$Department of Physics, Washington University, St. Louis, MO 63130, USA}\\
\normalsize{$^{2}$Department of Physics, Harvard University, Cambridge, MA 02138, USA}\\
\normalsize{$^{3}$Department of Physics, University of California, Berkeley, CA 94720, USA}\\
\normalsize{$^{4}$Institute of Materials Science and Engineering, Washington University, St. Louis, MO 63130, USA}\\
\normalsize{$^{5}$Materials Science Division, Lawrence Berkeley National Laboratory, Berkeley, CA 94720, USA}\\
\normalsize{$^*$These authors contributed equally to this work.}\\
\normalsize{$^\dag$To whom correspondence should be addressed; E-mail:  bingtian\_ye@berkeley.edu}\\
\normalsize{$^\ddag$To whom correspondence should be addressed; E-mail:  zu@wustl.edu}\\
}

\date{\today}

\maketitle

\tableofcontents

\section{Diamond sample}
The diamond sample (Element Six DNV-B14) in this work is a $3~\mathrm{mm}\times3~\mathrm{mm}\times0.5~\mathrm{mm}$ single crystal diamond grown by chemical vapor deposition (CVD). %
The sample is developed through a patented process with deliberate and controlled nitrogen-vacancy (NV) doping \cite{PurpleDiamond,DNVB14}, resulting an estimated total NV density $\rho_\mathrm{NV}\sim 4.5~$ppm quoted by the company.
We also independently characterize the NV density of the sample from the measured spin coherent dynamics and extract a density $(\sim 2.8\pm0.4)~$ppm, consistent with the reported value (see Section IV).

\section{Experimental setup}

We characterize the dynamics of NV centers using a home-built confocal laser microscope. A $532~$nm laser (Millennia eV High Power CW DPSS Laser) is used for both NV spin initialization and detection. The laser is shuttered by an acousto-optic modulator (AOM, G$\&$H AOMO 3110-120) in a double-pass configuration to achieve $>10^5:1$ on/off ratio. An objective lens (Olympus LCPlanFl 40x/0.60NA) focuses the laser beam to a diffraction limited spot with diameter $\sim 0.6~\mu$m, and collects the NV fluorescence. The fluorescence is then separated from the laser beam by a dichroic mirror, and filtered through a long-pass filter before being detected by an avalanche photodiode (Thorlabs). The signal is processed by a data acquisition device (National Instruments USB-6343). The objective lens is mounted on a piezo objective scanner (Physik Instrumente PD72Z1x PIFOC), which controls the position of the objective and scans the laser beam vertically. The lateral scanning is performed by an X-Y galvanometer (Thorlabs GVS212).

The NV centers are created randomly in the sample, so there exist 4 different crystalline orientations of the spin defects. To isolate one group of NV centers, we position a permanent magnet near the diamond to create an external magnetic field $\mathrm{B}\sim 350~$G along one of the NV axes. 
Under this magnetic field, the $|m_s = \pm1\rangle$ sublevels of the aligned NV group are separated due to Zeeman effect, and exhibits a splitting $2\gamma_e B$, where $\gamma_e = (2\pi)\times2.8~$MHz/G is the gyromagnetic ratio of the NV electronic spin.
A resonant microwave drive with frequency $(2\pi)\times 1.892~$GHz is applied to address the NV transition between $|m_s=0\rangle \Longleftrightarrow |m_s=-1\rangle$ sublevels and isolate an effective two-level system.
We note that at this magnetic field, after few microsecond of laser pumping, the associated spin-1 $^{14}$N nuclear spin of the NV center is highly polarized to $|m_I = +1\rangle$ via the excited state level anti-crossing (esLAC) \cite{jacques2009dynamic, fischer2013optical, zu2014experimental}.

The microwave driving field is generated by mixing the output from a microwave source (Stanford Research SG384) and an arbitrary wave generator (AWG, Chase Scientific Wavepond DAx22000). 
Specifically, a high-frequency signal at $(2\pi)\times 1.767~$GHz from the microwave source is combined with a $(2\pi)\times 0.125~$GHz signal from the AWG using a built-in in-phase/quadrature (IQ) modulator, so that the sum frequency at $(2\pi)\times 1.892~$GHz is resonant with the NV $|m_s=0\rangle \Longleftrightarrow |m_s=-1\rangle$ transition. 
By modulating the amplitude, duration and phase of the AWG output, we can control the strength, rotation angle and axis of the microwave pulses.
The microwave signal is amplified by a high-power amplifier (Mini-Circuits ZHL-15W-422-S+) and delivered to the diamond sample through a coplanar wave guide. The microwave is shuttered by a switch (Minicircuits ZASWA-2-50DRA+) to prevent any potential leakage. 
All equipment are gated through a programmable multi-channel pulse generator (SpinCore PulseBlasterESR-PRO 500) with $2$~ns temporal resolution.

We remark that in our experiment, the $\frac{\pi}{2}$- and $\pi$-pulse strength is set to $\Omega_\mathrm{p} = (2\pi)\times10~$MHz while the Floquet and quasi-Floquet driving fields strength is set to $\Omega = (2\pi)\times0.05~$MHz. 
Therefore, a good vertical resolution of the AWG is crucial to create pulses with high fidelity.
The AWG we use has a vertical resolution of 12-bit ($4096:1$) which is sufficient to generate high-quality microwave pulses for the experiment. 

\section{Differential measurement scheme}

The measurement sequence in this work is performed with a differential readout to mitigate the effect of NV charge dynamics under laser pumping and reliably extract the spin polarization difference between NV $\ket{m_s=0}$ and $\ket{m_s=-1}$ states (Fig.~\ref{fig:Dark}) \cite{mrozek2015longitudinal,choi2017depolarization, mittiga2018imaging,aslam2013photo,manson2018nv,jayakumar2016optical,bluvstein2019identifying, Zu2021emergent}.
In particular, we first let the NV charge dynamics reach a steady state by waiting for $0.5$~ms without any laser illumination (I).
After charge equilibration, we apply a $10~\mu$s laser pulse to initialize the NV centers (II).
After laser polarization, we apply first a global $\frac{\pi}{2}$ pulse along $\hat{y}$ direction to prepare the spin ensemble to a superposition state, followed by a periodic or quasi-periodic microwave driving field (III).
After the spin evolution, a final $\frac{\pi}{2}$ pulse along $-\hat{y}$ is used to rotate the spin state back to the $\hat{z}$ direction for spin polarization detection (IV).
By repeating the same procedure but with a final NV $\frac{\pi}{2}$-pulse along the positive $+\hat{y}$ axis before readout, we measure the fluorescence of an orthogonal spin state, and can use the difference between the two measurement to faithfully obtain the NV spin polarization \cite{choi2017depolarization,mrozek2015longitudinal}.
Since this general procedure is applied to all experiments in this work, in the main text we highlight the pulse sequences corresponding to the first II, III and IV regions.

\figSuppDark

\section{Dynamical decoupling sequence to cancel on-site field}

The presence of other spins in the diamond lattice (e.g. $^{13}$C nuclear spins and substitutional nitrogen spins) leads to an effective on-site random field at each NV center via the Ising portion of the dipolar interaction (see section IX). In experiments, we apply dynamical decoupling sequences to eliminate the effect of such static fields. After preparing the NV system into $\otimes_{i} \frac{|0\rangle_i+|-1\rangle_i}{\sqrt{2}}$ with a global $\frac{\pi}{2}$-pulse along $\hat{y}$ axis, we apply a series of $\pi$-pulses around $\pm \hat{x}$ axes, [$\tau/2$~---~$\pi_{\hat{x}}$~---~$\tau$~---~$\pi_{\hat{x}}$~---~$\tau$~---~$\pi_{-\hat{x}}$~---~$\tau$~---~$\pi_{-\hat{x}}$~---~$\tau/2$], with a fixed inter-pulse spacing $\tau = 0.1~\mu$s (Fig.~\ref{fig:SuppNVDensity}a top inset). 

In this case, the system dynamics is determined by the time evolution $U = [e^{-i\sum_{i} S^x_i \pi} e^{-i \mathcal{H} \tau}]^N$, where
\begin{equation} \label{eq1}
\begin{split}
\mathcal{H} = \mathcal{H}_0 + \sum_{i} h_i S^z_i = -\sum_{i<j} \frac{J_0 \mathcal{A}_{i,j}}{r^3_{i,j}}(S^z_i S^z_j - S^x_i S^x_j - S^y_i S^y_j) + \Omega \sum_{i} S^x_i + \sum_{i} h_i S^z_i,
\\
\end{split}
\end{equation}
is the Hamiltonian including the on-site random field, and $e^{-i\sum_{i} S^x_i \pi}$ represents the dynamical decoupling pulses, and $N$ is the number of cycles. The dynamical decoupling pulses can be well described by  $e^{-i\sum_{i} S^x_i \pi}$ because the driving strength is $(2\pi)\times 10~$MHz, more than two orders of magnitude larger than any other energy scale in the Hamiltonian $\mathcal{H}$. 
Since we choose the fixed inter-pulse spacing $\tau$, much smaller than any other timescales in the dynamics, the evolution of the system is well approximated by the unitary
\begin{equation} \label{eq1}
\begin{split}
U\approx [e^{-i\mathcal{H}_0\tau}]^N = e^{-i\mathcal{H}_0 N\tau} = e^{-i\mathcal{H}_0 t},
\end{split}
\end{equation}
where $t=N\tau$ is the total evolution time.
Practically, in any experimental implementation of decoupling sequence, the presence of pulse errors can potentially affect the dynamics and the conclusion. In order to rule out such effect, we implement the decoupling sequence with alternating phases along $\hat{x}$ and $-\hat{x}$ axes to compensate the error in the rotation angle. Additionally, we choose the $\pi$-pulses only along $\pm\hat{x}$ axes (rather than a conventional XY-$8$ sequence) so that the decoupling pulses do not alter the Hamiltonian of the driven NV system $\mathcal{H}_0$, but only cancel the on-site field disorder from the bath spins.

To characterize the NV density from experiment, we perfrom the $T_2^{XX}$ measurement and compare the measured decoherence timescale to numerical simulations. Figure~\ref{fig:SuppNVDensity}a shows the measured coherent decay of the NV centers with \emph{only} the dynamical decoupling sequence ($\tau = 0.5~\mu$s).
By fitting the data using a single exponential decay profile, we extract a coherence time, $T_2^\mathrm{XX} = (6.8\pm0.8)~\mu$s.
We sweep the value of inter-pulse spacing $\tau$ from $0.1~\mu$s to $1~\mu s$, and find that the corresponding decay timescales do not change significantly (Figure~\ref{fig:SuppNVDensity}a inset).
We attribute the slightly decrease of $T_2^\mathrm{XX}$ at smaller $\tau$ to the accumulated imperfections of the applied pulses. We emphasize that the measured $T_2^\mathrm{XX} = (6.8\pm0.8)~\mu$s should not be thought of as extrinsic decoherence, but rather as a consequence of the coherent dipolar interactions between NV centers.

The numerical simulation is performed using 1 central NV surrounded by 8 randomly positioned NV centers with varying spin concentration $\rho_{\frac{1}{4}}$.
Here the subscript $\frac{1}{4}$ represents a fraction of $\frac{1}{4}$ of the total NV centers in diamond, since under an applied large external magnetic field ($\sim 350~$G), only one of the four crystallographic axes of NV centers are being measured in experiment.
For each sampled density $\rho_{\frac{1}{4}}$, we average over 1000 realizations of positional disorder to get a smooth decoherent dynamics of the central NV center.
We extract the theoretical values of $T_2^\mathrm{XX}$ at different NV density by fitting the simulated profiles to a single exponential decay (Figure~\ref{fig:SuppNVDensity}b).
Comparing the experimental measured $T_2^\mathrm{XX}$ to the numerics, we estimate a spin density $\rho^\mathrm{exp}_{\frac{1}{4}}\approx(0.7\pm0.1)$~ppm.
This corresponds to a total NV concentration $\rho^\mathrm{exp} = 4\times\rho^\mathrm{exp}_{\frac{1}{4}} \approx (2.8\pm0.4)$~ppm, consistent with the quoted NV density $\rho_\mathrm{NV}\sim 4.5$~ppm from the company.

\figSuppNVDensity

\section{Extraction of the heating timescales}

In this section, we examine the robustness of the fitting scheme we apply to obtain the heating timescales $\tau^{*}$. 
In particular, through out the main text, we focus on the NV polarization dynamics after a transient behavior with $t\gtrsim100~\mu$s, to ensure that the system is already in a local equilibrium state. 
The heating timescale is then extracted by fitting the late-time dynamics to a single exponential decay, $\sim A e^{-(\frac{t}{\tau^*}+\frac{t}{T_0})}$, where $T_0 = (0.82 \pm 0.03)~$ms is independently characterized from the undriven case.
Below we provide additional data analysis by varying the fitting ranges and the fitting functional forms.
Crucially, we observe that these changes do not lead to qualitatively change of the measured heating timescales and thus our conclusions. 

\subsection{Varying fitting ranges}

To reliably extract the heating timescales, $\tau^{*}$, it is essential to isolate the late-time prethermal behaviors from the early-time initial equilibration.
Here, we show that when $t\gtrsim60~\mu$s, the NV polarization has already entered the prethermal regime, and the extracted scaling of heating time, $\tau^{*}$, with respect to driving frequency, $\omega$, is not sensitive to a specific starting point of the measured dynamics.
Fig.~\ref{fig:SuppFittingStartingPt}a depicts an example of the measured dynamics in the Floquet case with driving frequency $\omega = (2\pi)\times0.104~$MHz. Specifically, we fit the late-time data using a single exponential decay, and change the fitting range starting from $60~\mu$s, $80~\mu$s, $100~\mu$s and $120~\mu$s respectively. The corresponding fitted curves agree well with each other. In Fig.~\ref{fig:SuppFittingStartingPt}b, we plot the dependence of heating time, $\tau^{*}$, with driving frequency, $\omega$, for the four different fitting ranges. 
Crucially, in all four cases, we observe an exponential increment of the heating time $\tau^{*}$ with $\omega$ for more than an order of magnitude, demonstrating the existence of a parametrically long-lived prethermal regime in our long-range dipolar system.

\figSuppFittingStartingPt

\subsection{Varying fitting functional forms}

We also investigate the heating time by fitting the measured dynamics to a stretched exponential decay profile, $\sim A e^{-(\frac{t}{\tau^*})^{0.5}} e^{-\frac{t}{T_0}}$.  
Such functional form has been recently used to capture the measured quantum dynamics of a solid-state spin system with strong positional disorder~\cite{choi2017depolarization,Peng2021Prethermal, Zu2021emergent,davis2021probing}.
Figure~\ref{fig:SuppFittingForm}a shows an example of the two fitting curves using single exponential profile and stretched exponential profile respectively [$\omega = (2\pi)\times0.104~$MHz]. 
Crucially, we find that both functional forms capture the measured late-time NV spin dynamics equally well, presumably because there is not enough early-time data to distinguish between the two decay forms.
The extracted heating time, $\tau^{*}$, exhibits a small difference in the two cases, as shown in Figure~\ref{fig:SuppFittingForm}b.
Nevertheless, in either case, we observe an exponential increment of $\tau^{*}$ with driving frequency $\omega$, confirming the existence of prethermalization in our dipolar NV ensemble.

\figSuppFittingForm

\subsection{Distinguishing exponential dependence of heating time from power-law scaling}
In this section, we provide additional analysis of the experimentally measured heating time, $\tau^*$, to distinguish the observed exponentially slow heating based on Floquet prethermalization from conventional power-law scaling.
In particular, we also try to fit the heating time using a second order polynomial form, $\tau^* = A\omega^2$, as expected from a standard perturbation theory.
Yet, as shown in Figure~\ref{fig:ExpFitting}(a),the second order polynomial form (dashed red line) hardly reproduces our experimental result.
In Figure~\ref{fig:ExpFitting}(b), we plot 
the relative residues from both exponential (blue) and second order polynomial (red) fittings, and find that the second order polynomial fitting results in nearly an order of magnitude larger $\chi^2$ value.
Therefore, we believe our experiment is more consistent with the exponential scaling predicted by Floquet prethermalization rather than second order power-law scaling predicted from standard perturbation theory.

We remark that it is extremely difficult to unequivocally demonstrate a difference between an exponential scaling and a \emph{generic} power-law scaling, $\tau^*\sim\omega^{\beta}$.
Specifically, one can always choose a very large power $\beta$, so that the power-law function can capture the real exponential scaling within a given range of $\omega$.
For instance, in Figure~\ref{fig:ExpFitting}, we also fit our experimental data using a fourth-order polynomial, $\tau^* = A \omega^4$.
We find that the exponential (blue) and the fourth order polynomial (yellow) resembles each other. Along with the $\chi^2$ test, the result indicates that the exponential form and the fourth order polynomial describe our experimental data equally well. 
However, we note that the seemingly robust fourth order polynomial scaling falls short of a simple physical interpretation from perturbation theory.

\figExpFitting

\section{Micromotion in quasi-periodically driven systems}

In a conventional Floquet system, if one does not measure the spin dynamics at integer multiples of the Floquet period (i.e. stroboscopic time), the time-dependent part
of the Hamiltonian will not average to zero, thus leads to a small periodic oscillation on top of the measured spin dynamics, known as micromotion \cite{sambe1973steady, goldman2014periodically, eckardt2015high}.
Figure~\ref{fig:Oscillation} shows an example of the measured Floquet micromotion in experiment if one intentionally measures at non-stroboscopic times.
With a single driving frequency, $\omega = (2\pi)\times0.088~$MHz, the observed micromotion exhibits a period of $2\pi/\omega \approx 11.4~\mu$s, which describes the time evolution of the system within one Floquet cycle.
\figOscillation

However, in a quasi-periodic driven system with two incommensurate driving frequencies, $\omega$ and $\varphi \omega$, there does not exist a stroboscopic time to be an integer multiple of both driving periods. 
As a result, the measured late-time spin polarization displays a quasi-periodic micromotion on top of the perthermal dynamics (Fig.~2d in main text).
We find that the relative amplitude of the micromotion, $W$, which can be calculated using the standard deviation of the spin polarization divided by the average, scales quadratically with the inverse of driving frequency, $W \sim \omega^{-2}$.
To reliably extract the heating time, $\tau^{*}$, we apply a rolling average method on the experimental data to obtain a smooth profile of the spin dynamics (Figure~\ref{fig:RollingAverage}).
Specifically, we first choose a time at the integer multiples of one driving period, $t_c =\frac{2\pi N}{\omega}$, and measure four data points at times $t = [t_c - \frac{3}{8} \frac{2\pi}{\varphi\omega}, t_c - \frac{1}{8} \frac{2\pi}{\varphi\omega}, t_c + \frac{1}{8} \frac{2\pi}{\varphi\omega}, t_c + \frac{3}{8} \frac{2\pi}{\varphi\omega}]$.
We then take the average value of the four points, and use it to represent the spin polarization at time $t_c$.
After the rolling average, the measured spin dynamics display smooth decay profiles (main text Fig.~2e) which can then be used to obtain the heating times in quasi-periodic driven systems.

\figRollingAverage

\section{Two-dimensional Fourier spectrum of the quasi-Floquet drive}

In this section, we give a detailed explanation of the two-dimensional Fourier spectrum of the quasi-periodic drive shown in main text Figure~3c.
Specifically, we can express the quasi-Floquet drive $f(t)$ using a two-dimensional Fourier space,
\begin{equation}
f(t) = \sum_{n_1,n_2 = 0, \pm1, ...} F_{n_1, n_2}  e^{i (n_1 \omega + n_2\varphi \omega) t},
\end{equation}
where $F_{n_1,n_2}$ characterizes the Fourier component of the driving field at frequency $\nu = (n_1 \omega + n_2\varphi \omega)$ (green dots in main text Figure~3c).
By projecting the two-dimensional $F_{n_1,n_2}$ onto a one-dimensional line with slope $\varphi$, we obtain the Fourier spectrum $F(\nu)$ of the quasi-periodic drive in main text Figure~3c.

A few remarks are in order. 
First, unlike the conventional Floquet prethermalization, in quasi-periodic driven systems, even with large driving frequencies [$\omega, \varphi\omega \gg \widetilde{J}$], there can exist $\nu = (n_1 \omega + n_2\varphi \omega)<\widetilde{J}$ with non-zero $F(\nu) = F_{n_1, n_2}$, which leads to potential energy absorption.
Second, for any given driving frequency $\omega$, only when $n_1, n_2 \gtrsim \omega/\widetilde{J}$, one can find $\nu = n_1\omega+n_2\varphi\omega<\widetilde{J}$.
As a result, as long as $F_{n_1,n_2}$ is sufficiently small with increasing $n_1$ and $n_2$ (in other words, $f(t)$ is sufficiently smooth in time), the energy absorption from the multi-photon processes is suppressed, enabling the existence of a long-lived prethermal regime in quasi-periodic driven systems \cite{Else2020quasi}.

\section{Local energy scale}
\subsection{Summary of existing results in different parameter regimes}
In a long-range power-law interacting periodically or quasi-periodically driven system, the existence of prethermalization highly depends on the interplay between the dimensionality and the interaction range. 
In this section, we summarize both the theoretical and experimental results, in order to make it clear how our work is related with and adds to the previous literature (Table~\ref{tab:prethermal_req}). 

The prethermalization phenomenology consists of two ingredients: 1) a (stretched) exponentially slow heating time, and 2) an effective static Hamiltonian governing the prethermal dynamics.
To be concrete, we consider a $D$-dimensional system with a power-law interaction, $J(r)\sim \mathcal{A}/r^\alpha$, where prefactor $\mathcal{A}$ characterizes the possible angular dependence of the interaction. 
As explained in the main text and detailed below, the existence of exponentially slow heating relies on the local energy scale $\widetilde{J} \approx [\int J(r)^2\rho d^D \textbf{r}]^{\frac{1}{2}}$ being finite in the presence of disorder. 
Crucially, when the angular dependence averages to zero [$\bar{J}(r) = 0$], a finite $\widetilde{J}$ requires $\alpha>D/2$; in the opposite case [$\bar{J}(r) \neq 0$], $\alpha>D$ is required. 
Moreover, we remark that the presence of exponentially slow heating does not guarantee the existence of a static Hamiltonian, $\mathcal{H}_\mathrm{eff}$, to effectively generate the dynamics in the prethermal regime. 
On the one hand, a strict analytical proof for such prethermal Hamiltonian relies on the results of Lieb-Robinson bounds, which constrain the speed of spreading of local perturbations \cite{kuwahara2016floquet,machado2020long}. 
On the other hand, even in systems without Lieb-Robinson bounds, numerical evidences show that a formally constructed effective Hamiltonian by Floquet-Magnus expansion can still describe the prethermal dynamics well \cite{machado2019exponentially,machado2020long,ye2021floquet}. 

Our three-dimensional NV system with dipolar interaction ($\alpha=3$) sits in the regime of $D/2<\alpha<D$ and $\bar{J}(r)=0$, where there are not many results from the previous literature, especially on the experimental front. 
Looking forward, the recently developed two-dimensional NV systems can enable the experimental investigation of prethermal behaviors in $D<\alpha<2D$ regime \cite{davis2021probing}. 

\begin{table*}[h!]
  \renewcommand{\arraystretch}{1.5}
 \begin{tabular}{c|c||c|c} 
    \hline \hline
   \multicolumn{2}{c||}{\makecell{condition}} &  exponentially slow heating & effective prethermal Hamiltonian \\ \hline \hline
    \multicolumn{2}{c||}{\makecell{$\alpha\leq D/2$}} & \ding{55} & \ding{55} \\ \hline
    \multirow{2}{*}{$D/2<\alpha\leq D$}&~~$\bar{J}(r)\neq 0$~~~& \ding{55} & \ding{55} \\\cline{2-4} &~~$\bar{J}(r)= 0$~~~&\makecell{
    analytical proof \cite{ho2018bounds}\\ experimental evidence [\emph{this work}]} & \makecell{lack of analytical proof\\ experimental evidence [\emph{this work}]} \\ \hline
    \multicolumn{2}{c||}{\makecell{$D<\alpha\leq 2D$}} & \makecell{
    analytical proof \cite{kuwahara2016floquet,mori2016rigorous,abanin2017rigorous}\\ lack of experimental evidence} & \makecell{
    lack of analytical proof\\numerical evidence \cite{machado2019exponentially}\\ lack of experimental evidence} \\\hline
    \multicolumn{2}{c||}{\makecell{$\alpha>2D$\\short-range}} & \makecell{
    analytical proof \cite{kuwahara2016floquet,mori2016rigorous,abanin2017rigorous}\\ experimental evidence \cite{Peng2021Prethermal, rubio2020floquet}} & \makecell{
    analytical proof \cite{kuwahara2016floquet,mori2016rigorous,abanin2017rigorous}\\ experimental evidence \cite{Peng2021Prethermal, rubio2020floquet}} \\
    \hline \hline 
  \end{tabular}
  \caption{
  Summary of the results on prethermalization in the literature.
    }
     \label{tab:prethermal_req}
\end{table*}

\subsection{Key idea to estimate local energy scale}
Let us start with the physical intuition of the local energy scale in our long-range dipolar interacting system with angular dependence. The interactions can be both positive and negative which accords a cancellation of many terms in the response function of the system. Crucially, the angular dependence of the dipolar interaction leads to an exact cancellation of the positive and the negative interaction strength to the leading order. Therefore, to estimate the local energy scale in the response function, one needs to go to at least the second order, i.e. the variance of the interaction.

To be more concrete, we also sketch the essence in the mathematical proof of the estimation of local energy scale to ensure that the major claims of this Letter is self-contained, as a strict analytical study of the prethermalization with the presence of disorder can be found in the previous literature\cite{ho2018bounds}, which further clarify how the lowest order of the interaction strength gets averaged out.

Let us consider a generic model with both long-range two-body interaction and single-body terms, of which the Hamiltonian is written as: 
\begin{equation}
H = \sum_{ij}J_{ij} \hat{O}_{ij}+\sum_i J_i \hat{O}_i, 
\end{equation}
where $\hat{O}_{ij}$ ($\hat{O}_i$) are two-body (single-body) operators, and $J_{ij}$ ($J_i$) are the corresponding coupling strengths which varies in time with a frequency of $\omega$. 
In the high-frequency regimes, considering the high-order processes, the heating effect is induced by the transition rate between different energy levels and is thus bounded by the high-order commutators \cite{kuwahara2016floquet,mori2016rigorous,abanin2017rigorous,ho2018bounds}
\begin{equation}
    \frac{1}{\omega^p}\langle J_{\mu_1}J_{\mu_2}J_{\mu_3}J_{\mu_4}\cdots\rangle\mathrm{Tr}(\hat{O}_{\mu_1}[[[\hat{O}_{\mu_2},\hat{O}_{\mu_3}],\hat{O}_{\mu_4}],\cdots]^{(p)}), 
\label{tab:J_average}
\end{equation}
where the superscript $(p)$ denotes that there are $p$ layers of commutator in the expression, the subscript $\mu$ denotes either the two indices associated with a two-body term or the index associated with a single-body term, and the bracket $\langle\cdots\rangle$ denotes the average over positional configurations of the spins. 
In generic short-range systems, all the $\langle J_{\mu_i}\rangle$ remains finite and non-zero, so Eq.~S5 $\sim p!(\frac{J}{\omega})^p$, where $J = \langle J_{\mu_i}\rangle$ is considered as the local energy scale. 
One can eventually prove an exponentially small heating rate by finding the optimal $p\sim \omega$\cite{kuwahara2016floquet,mori2016rigorous,abanin2017rigorous,ho2018bounds}. 
In contrast, for each two-body interaction $J_{\mu_i}$, if $\langle J_{\mu_i}\rangle=0$, then it has to appear at least twice for non-zero $\langle J_{\mu_1}J_{\mu_2}J_{\mu_3}J_{\mu_4}\cdots\rangle$. 
Therefore, the size of Eq.~S5 should be estimated as
\begin{equation}
    \sim p!\frac{\langle J_{\mu}^2\rangle^{\frac{p}{2}}}{\omega^p}. 
\end{equation}
Let us note that such estimation of the local energy scale only accounts the terms in Eq.~S5, in which only two-body terms are involved. 
However, one should also, in principle, consider single-body terms, with the requirement that each two-body $J_{\mu_i}$ term has to appear twice still being satisfied. 
Nevertheless, as long as the local energy scale of the two-body terms $\langle J_{\mu}^2\rangle^{\frac{1}{2}}$ is of the same order of the single-body field, we can use their values as a typical local energy scale to estimate the heating rate. Going back to our dipolar interaction in experiment, $\langle J_\mu^2 \rangle ^\frac{1}{2} = \{\int [\frac{J_0}{r^3} (3 \cos^2{\theta}-1)]^2~\rho r^2 \sin{\theta}~dr~d\theta~d\phi)\}^\frac{1}{2} = \sqrt{\frac{16\pi}{15}} J_0\rho$.

\section{Relation between Floquet prethermalization and Rotating Wave Approximation}

In this section, we clarify the relationship between our current work on Floquet prethermalization and related works in the NMR literature. 
The rotating wave approximation (RWA),
which is widely used in the context of NMR systems (and in almost all AMO systems), is essentially an application of Floquet engineering. 
However, one key difference between the RWA and Floquet prethermalization is the parametric
timescale for the approximation to be valid. 
In particular, RWA utilizes energy conservation arguments to eliminate the leading
order correction from the periodic drive; in experiments, this can lead to relatively long time-scales; but crucially, from a parametric perspective, the RWA is only valid for a time that scales polynomially in the frequency of the drive.
In contrast, Floquet prethermalization analyzes the convergence of a Magnus expansion to show that an effective Hamiltonian governing the dynamics can
emerge for intermediate time-scales. 
Crucially, unlike the RWA case, numerical
and analytic results demonstrate that Floquet prethermalization persists for an
exponentially long time in the driving frequency.

To be specific, the validity of RWA relies on a separation between different energy scales. As an example, for the NV center in diamond, the spin transition frequency is on the order of gigahertz while the interactions are on the order of megahertz. Similarly, for NMR systems, the nuclear spin transition frequency can be on the order of megahertz while interaction strengths are often
on the order of kilohertz; this causes the system dynamics to become governed by energy conserving terms (until late times). We note that, again in contrast to Floquet prethermalization, the above arguments do not rely on any details of
the driven many-body systems, including the interaction range, dimensionality,
integrability of the system, and the types of Floquet drive (single v.s. multi frequencies, smooth v.s. sharp amplitude modulation), etc.

Previous NMR/AMO
experiments using the rotating wave approximation can be perhaps be recast
using the language of Floquet prethermalization. The challenge is that since the
driving frequencies in those experiments were typically much larger than the
other energy scales of the system, it becomes hard to measure to late enough
times to distinguish between the polynomially long time-scale predicted by RWA
argument and the exponentially long time-scale predicted by Floquet prethermalization. 
Perhaps most importantly, these prior experiments do not focus on
exploring the parametric dependence of the thermalization time-scale with the
driving frequency, a key observable identify Floquet prethermal behavior.
On the other hand, our work, and more broadly the recent resurgence on Floquet prethermalization, precisely explore the response of the many-body system as the driving frequency is changed, with the goal of exploring the minimal requirements for
Floquet prethermalization.

\section{Dipolar Hamiltonian under the rotating-wave approximation}

In this section, we derive the dipolar interacting Hamiltonian of the NV ensemble described by Eq.~(1) from the main text. In the laboratory frame, the spin dipole-dipole interaction between two NV defects can be written as:
\begin{equation} \label{eq1}
\mathcal{H}_{dip} 
= - \frac{J_0}{r^3}[3(\hat{\mathcal{S}}_1\cdot\hat{n})(\hat{\mathcal{S}}_2\cdot\hat{n})-\hat{\mathcal{S}}_1\cdot\hat{\mathcal{S}}_2]    ,
\end{equation}
where $J_0 = (2\pi)\times52~$MHz$\cdot$nm$^3$, $r$ and $\hat{n}$ denote the distance and direction unit vector between two NV centers, and $\hat{\mathcal{S}}_1$ and $\hat{\mathcal{S}}_2$ are the NV spin-$1$ operators. Our experiments only focus on an effective two-level system $\{|m_s = 0\rangle, |m_s = -1\rangle\}$, so the spin operators in the restricted Hilbert space are:
\begin{equation} \label{eq}
\begin{split}
\mathcal{S}^z =
\begin{bmatrix}
0 & 0 \\
0 & -1
\end{bmatrix}	,~
\mathcal{S}^x = \frac{1}{\sqrt{2}}
\begin{bmatrix}
0 & 1 \\
1 & 0
\end{bmatrix}	,~
\mathcal{S}^y = \frac{1}{\sqrt{2}}
\begin{bmatrix}
0 & -i \\
i & 0
\end{bmatrix}   .
\end{split}
\end{equation}
Also, we can define the spin raising and lowering operators:
\begin{equation} \label{eq}
\mathcal{S}^+ =
\begin{bmatrix}
0 & 1 \\
0 & 0
\end{bmatrix}
= \frac{\mathcal{S}^x + i\mathcal{S}^y}{\sqrt{2}} ,~
\mathcal{S}^- =
\begin{bmatrix}
0 & 0 \\
1 & 0
\end{bmatrix}
= \frac{\mathcal{S}^x - i\mathcal{S}^y}{\sqrt{2}} ,
\end{equation}
and rewrite spin operators in terms of the raising and lowering operators:
\begin{equation} \label{eq}
\mathcal{S}^x = \frac{\mathcal{S}^+ + \mathcal{S}^-}{\sqrt{2}} ,~
\mathcal{S}^y = \frac{\mathcal{S}^+ - \mathcal{S}^-}{i\sqrt{2}} .
\end{equation}
Then we can expend the dipolar interaction in Eq.~(S1) as:
\begin{equation} \label{eq2}
\begin{split}
\mathcal{H}_{dip} = - \frac{J_0}{r^3}\times
&\left\{3
 \left[\mathcal{S}^z_1 n_z + \frac{(\mathcal{S}^+_1 + \mathcal{S}^-_1)n_x}{\sqrt{2}} +
 \frac{(\mathcal{S}^+_1 - \mathcal{S}^-_1)n_y}{i\sqrt{2}}\right]
 \left[\mathcal{S}^z_2 n_z + \frac{(\mathcal{S}^+_2 + \mathcal{S}^-_2)n_x}{\sqrt{2}} +
 \frac{(\mathcal{S}^+_2 - \mathcal{S}^-_2)n_y}{i\sqrt{2}}\right] \right.\\
&- \left.\left[\mathcal{S}^z_1 \mathcal{S}^z_2 +
 \frac{(\mathcal{S}^+_1 + \mathcal{S}^-_1)}{\sqrt{2}}
 \frac{(\mathcal{S}^+_2 + \mathcal{S}^-_2)}{\sqrt{2}} + 
 \frac{(\mathcal{S}^+_1 - \mathcal{S}^-_1)}{i\sqrt{2}}
 \frac{(\mathcal{S}^+_1 - \mathcal{S}^-_1)}{i\sqrt{2}}\right]\right\} .
\end{split}
\end{equation}

For each NV center, there is a splitting $\Delta=(2\pi)\times1.892~$GHz between the two levels $|m_s = 0\rangle$ and $|m_s = -1\rangle$ along the $z$ direction (under a external magnetic field $\sim 350$~G). Therefore, the evolution driven by $\Delta\mathcal{S}^z$ is worth to be noted. Consider a quantum state $|\phi\rangle$ in the rotating frame $|\varphi\rangle = e^{-i\Delta \mathcal{S}^z t}|\phi\rangle$. If we apply Schrödinger equation:
\begin{equation} \label{eq}
\begin{aligned}
i\partial_{t}|\varphi\rangle
 &= (\Delta\mathcal{S}^z + \mathcal{H}_{dip})|\varphi\rangle\\
i\partial_{t}(e^{-i\Delta\mathcal{S}^z t}|\phi\rangle)
 &= (\Delta\mathcal{S}^z + \mathcal{H}_{dip})(e^{-i\Delta\mathcal{S}^z t}|\phi\rangle)\\
\Delta\mathcal{S}^z e^{-i\Delta\mathcal{S}^z t}|\phi\rangle + e^{-i\Delta\mathcal{S}^z t}i\partial_{t}|\phi\rangle
 &= \Delta\mathcal{S}^z e^{-i\Delta\mathcal{S}^z t}|\phi\rangle + \mathcal{H}_{dip} e^{-i\Delta\mathcal{S}^z t}|\phi\rangle\\
i\partial_{t}|\phi\rangle
 &= e^{i\Delta\mathcal{S}^z t}\mathcal{H}_{dip}e^{-i\Delta\mathcal{S}^z t}|\phi\rangle .
\end{aligned}
\end{equation}
Then we can define dipolar interaction Hamiltonian in rotating frame:
\begin{equation} \label{eq}
\tilde{\mathcal{H}}_{dip} = e^{i\Delta\mathcal{S}^z t} \cdot \mathcal{H}_{dip} \cdot e^{-i\Delta\mathcal{S}^z t} ,
\end{equation}
and the spin operators in the rotating frame:
\begin{equation} \label{eq}
\begin{aligned}
\tilde{\mathcal{S}}^z 
&= e^{i\Delta \mathcal{S}^z t} \cdot {\mathcal{S}^z} \cdot e^{-i\Delta \mathcal{S}^z t} = \mathcal{S}^z\\
\tilde{\mathcal{S}}^+ 
&= e^{i\Delta \mathcal{S}^z t} \cdot {\mathcal{S}^+} \cdot e^{-i\Delta \mathcal{S}^z t} = \mathcal{S}^+ \cdot e^{+i\Delta t}\\
\tilde{\mathcal{S}}^- 
&= e^{i\Delta \mathcal{S}^z t} \cdot {\mathcal{S}^-} \cdot e^{-i\Delta \mathcal{S}^z t} = \mathcal{S}^- \cdot e^{-i\Delta t} .
\end{aligned}
\end{equation}
In the rotating frame, rewrite the dipolar interaction Hamiltonian (Eq.S5):
\begin{equation} \label{eq3}
\begin{split}
\tilde{\mathcal{H}}_{dip} = - \frac{J_0}{r^3}\times
&\left\{3
 \left[\tilde{\mathcal{S}}^z_1 n_z + \frac{(\tilde{\mathcal{S}}^+_1 + \tilde{\mathcal{S}}^-_1)n_x}{\sqrt{2}} + \frac{(\tilde{\mathcal{S}}^+_1 - \tilde{\mathcal{S}}^-_1)n_y}{i\sqrt{2}}\right] 
 \left[\tilde{\mathcal{S}}^z_2 n_z + \frac{(\tilde{\mathcal{S}}^+_2 + \tilde{\mathcal{S}}^-_2)n_x}{\sqrt{2}} + \frac{(\tilde{\mathcal{S}}^+_2 - \tilde{\mathcal{S}}^-_2)n_y}{i\sqrt{2}}\right] \right.\\
&- \left.\left[\tilde{\mathcal{S}}^z_1 \tilde{\mathcal{S}}^z_2 +
 \frac{(\tilde{\mathcal{S}}^+_1 + \tilde{\mathcal{S}}^-_1)}{\sqrt{2}} \frac{(\tilde{\mathcal{S}}^+_2 + \tilde{\mathcal{S}}^-_2)}{\sqrt{2}} +
 \frac{(\tilde{\mathcal{S}}^+_1 - \tilde{\mathcal{S}}^-_1)}{i\sqrt{2}} \frac{(\tilde{\mathcal{S}}^+_2 - \tilde{\mathcal{S}}^-_2)}{i\sqrt{2}}\right]\right\} ,
\end{split}
\end{equation}
which can be simplified to
\begin{equation} \label{eq4}
\begin{split}
\tilde{\mathcal{H}}_{dip} = - \frac{J_0}{r^3}\times
&\left\{\right.(3n^2_z -1)\mathcal{S}^z_1 \mathcal{S}^z_2 + (\mathcal{S}^+_1 \mathcal{S}^-_2 + \mathcal{S}^-_1 \mathcal{S}^+_2)\left[\frac{3}{2}(n^2_x + n^2_y)-1\right]\\
&+ \frac{3}{2}\mathcal{S}^+_1 \mathcal{S}^+_2 e^{+2i\Delta t}(n^2_x - n^2_y - 2in_x n_y)
 + \frac{3}{2}\mathcal{S}^-_1 \mathcal{S}^-_2 e^{-2i\Delta t}(n^2_x - n^2_y + 2in_x n_y)\\
&+ 3\mathcal{S}^z_1 n_z \left[
    \frac{n_x}{\sqrt{2}}(\mathcal{S}^+_2 e^{+i\Delta t} + \mathcal{S}^-_2 e^{-i\Delta t}) +
    \frac{n_y}{i\sqrt{2}}(\mathcal{S}^+_2 e^{+i\Delta t} - \mathcal{S}^-_2 e^{-i\Delta t}) \right]\\
&+ 3\mathcal{S}^z_2 n_z \left[
    \frac{n_x}{\sqrt{2}}(\mathcal{S}^+_1 e^{+i\Delta t} + \mathcal{S}^-_1 e^{-i\Delta t}) +
    \frac{n_y}{i\sqrt{2}}(\mathcal{S}^+_1 e^{+i\Delta t} - \mathcal{S}^-_1 e^{-i\Delta t}) \right]\} .
\end{split}
\end{equation}
Since we are interested in spin-spin interaction dynamics with energy scale $J_0/r^3 \approx (2\pi)\times 0.01$~MHz that is much smaller than the splitting $\Delta \approx (2\pi)\times1.892$ GHz, we are able to drop the last six time-dependent terms and only keep the energy-conserving terms under the rotating-wave approximation. Additionally, considering $n^2_x + n^2_y + n^2_z = 1$, we get
\begin{equation} \label{eq5}
\begin{split}
\tilde{\mathcal{H}}_{dip} 
&= - \frac{J_0}{r^3}\times(3n^2_z -1)[\mathcal{S}^z_1 \mathcal{S}^z_2 - \frac{1}{2}\mathcal{S}^+_1 \mathcal{S}^-_2 - \frac{1}{2}\mathcal{S}^-_1 \mathcal{S}^+_2]\\
&= - \frac{J_0}{r^3}\times\frac{(3n^2_z -1)}{2}[2\mathcal{S}^z_1 \mathcal{S}^z_2 - \mathcal{S}^x_1 \mathcal{S}^x_2 - \mathcal{S}^y_1 \mathcal{S}^y_2] .
\end{split}
\end{equation}
We can rewrite the interacting Hamiltonian using normal spin-$\frac{1}{2}$ operators
\begin{equation} \label{eq}
\begin{split}
S^z = \frac{1}{2}
\begin{bmatrix}
1 & 0 \\
0 & -1
\end{bmatrix}	,~
S^x = \frac{1}{2}
\begin{bmatrix}
0 & 1 \\
1 & 0
\end{bmatrix}	,~
S^y = \frac{1}{2}
\begin{bmatrix}
0 & -i \\
i & 0
\end{bmatrix}.
\end{split}
\end{equation}
Specifically, we convert the effective two-level spin-$1$ operators to spin-$\frac{1}{2}$ operators, $\mathcal{S}^{x} = \sqrt{2}S^x$, $\mathcal{S}^{y} = \sqrt{2}S^y$, $\mathcal{S}^{z} = S^z+1/2$, and plug them into Eq.~(S11), 
\begin{equation} \label{eq6}
\mathcal{H}_{dip} = - \frac{J_0 \mathcal{A}}{r^3}(S^z_1 S^z_2 - S^x_1 S^x_2 - S^y_1 S^y_2) ,
\end{equation}
where $\mathcal{A} = 3n^2_z - 1$ is the angular dependent factor.

To derive the dipolar Hamiltonian of the entire NV spin ensemble, we simply sum up the interactions between every pair of NV spins:
\begin{equation} \label{eq7}
\mathcal{H}_{dip} = -\sum_{i<j} \frac{J_0 \mathcal{A}_{i,j}}{r^3_{i,j}}(S^z_i S^z_j - S^x_i S^x_j - S^y_i S^y_j),
\end{equation}
where $\mathcal{A}_{i,j}$ and $r_{i,j}$ represent the angular dependence of the long-range dipolar interaction and the distance between the $i^{th}$ and $j^{th}$ NV centers.

We remark that due to random positions of the NV centers in diamond lattice, the angular dependence term averages to zero.
This can be easily seen by writing the angular dependence term in spherical coordinate, $\mathcal{A}_{i,j} = 3n^2_z - 1 = 3\cos^2{\theta}-1$, and integrate it on the unit sphere over polar angle $\theta$ and azimuthal angle $\phi$,
\begin{equation} \label{eq}
\begin{split}
\overline{\mathcal{A}}_{i,j} 
= \frac{\int_{0}^{2\pi} d\phi \int_{0}^{\pi} \sin{\theta} d\theta \cdot
  (3\cos^2{\theta}-1)}{\int_{0}^{2\pi} d\phi \int_{0}^{\pi} \sin{\theta} d\theta}
= \frac{1}{2} \int_{0}^{\pi} \sin{\theta} d\theta \cdot (3\cos^2{\theta}-1)
= 0
\end{split}
\end{equation}

\section{Error estimation and propagation}
The errorbars for the measured spin polarization $\langle S^x(t)\rangle$ represent one standard deviation of the statistical error, and can be calculated via propagating the statistical errors from the measured NV photoluminescent counts. 
The errorbars on the measured NV counts are directly estimated assuming a Gaussian distribution with a mean value $N$ and a standard deviation $\sqrt{N}$, where $N$ is the detected photon number.
We note that in our figures, some errorbars are smaller than the data markers. The errorbars for the extracted heating time $\tau^*$ and prethermal plateau value $A(\omega)$ represent one standard deviation from the fitting.

\bibliography{Supp.bib}
\newpage